\documentclass[titlepage,12pt]{article}
\begin{document}

\begin{titlepage}
\thispagestyle{empty}
\begin{center}
{\Large\bf Evolution in Time of Moving Unstable Systems}\\[0.5cm]
{\large\bf M. I. Shirokov } 
\footnote{Bogoliubov Laboratory of Theoretical Physics,\\
Joint Institute for Nuclear Research,\\
141980 Dubna, Russia;\\
e-mail: shirokov@thsun1.jinr.ru}
\end{center}

\vspace*{1cm} \noindent
{\large\bf Abstract}\\
\noindent
Relativistic quantum theory shows that the known Einstein time dilation (ED)
approximately holds for the decay law of the unstable particle having definite
momentum $p$ (DP). I use a different definition of the moving particle as the
state with definite velocity $v$ (DV). It is shown that in this case the decay
law is not dilated. On the contrary, it is contracted as compared with the
decay law of the particle at rest. It is demonstrated that ED fails in both
DP and DV cases for time evolution of the simple unstable system of the
kind of oscillating neutrino. Experiments are known which show that ED
holds for mesons. The used theory may explain the fact by supposing that the
measured mesons are in DP state.

\vspace{1.5cm}
\begin{tabular}{cp{9cm}}
{\large\bf KEY WORDS:} & unstable particle; decay law; Einstein time dilation;
relativistic quantum mechanics; neutrino oscillation.
\end{tabular}

\end{titlepage}


\section{INTRODUCTION}
\label{intro}

Experimenters showed that the lifetime $\tau$ of a uniformly moving
unstable particle is equal to $\tau _0 \gamma$,
where $\tau _0$ is the lifetime of the particle at rest and $\gamma$
is the Lorentz factor $\gamma = (1-v^2/c^2)^{-1/2}$, e.g., see (Baily, 1977;
Farley, 1992). In other words, if $F(t)$ is the decay law of the unstable
particle moving in the laboratory frame and $F_0(t)$ is the decay law of
the same particle at rest, then $F_0(t)=\exp (-t/\tau_0)$
and $F(t)=\exp (-t/\tau_0\gamma)$ or
\begin{equation}
\label{eq1}
F(t)=F_0(t/\gamma)\,.
\end{equation}
A usual explanation of the fact is
based on the Einstein special theory of relativity and I call (\ref{eq1})
the Einstein dilation (ED). For example, M{\o}ller (1972) sets it forth as
follows:
\begin{quotation}
``In view of the fact that an arbitrary physical system can be
used as a clock, we see that any physical system which is moving
relative to a system of inertia must have a slower course of
development than the same system at rest. Consider for instance a
radioactive process. The mean life $\tau$ of the radioactive
substance, when moving with a velocity $v$, will thus be larger
than the mean life $\tau _0$ when the substance is at rest. From
(2.36) we obtain immediately $\tau = (1-v^2/c^2)^{-1/2}\tau _0$."
\end{quotation}

This argumentation may be complemented by the following possible
definition of the unit of time provided by radioactive substance:
this is the time interval during which the amount of the
substance decreases twice, e.g.

However, the standard clocks of the relativity theory are used
when obtaining Eq.~(2.36)
$$\Delta t=t_2-t_1=\gamma (t_2'-t_1')=(1-v^2/c^2)^{-1/2}\Delta\tau$$
which M{\o}ller mentions. He begins the derivation of this equation
with the phrase:
\begin{quotation}
``Consider a standard clock $C'$ which is placed at rest in $S'$
at a point on the $x'$-axis with the coordinate $x'=x_1'$."
\end{quotation}

However, such a quantum clock as an unstable particle cannot be at rest
(i.e., have zero velocity or zero momentum) and simultaneously be
at a definite point (due to the quantum uncertainty relation). So
the standard derivation of the moving clock dilation is
inapplicable for the quantum clock.

Another way of theoretical derivation may be used: to find the relativistic
quantum decay law $F(t)$ of the moving particle and to compare it with
the decay law $F_0(t)$ of the particle at rest. Lorentz transformation
of the space-time coordinates from one inertial frame to another is
not needed as well as the space coordinates themselves. The approach was
employed by (Exner, 1983; Stefanovich, 1996; Khalfin, 1997; Shirokov, 2004).
In these papers (below I shall refer to them as (ESKS)) the state of the
moving unstable particle was described by the eigenvector $\Psi_p$
of the momentum operator $\widehat{\vec{P}}$ (Exner (1983) used a packet with
almost exact momentum). One may state that the obtained decay law
$F_p(t)$ is consistent with ED, Eq.~(\ref{eq1}), see Sect.~3 below.
I use in Sect.~2 another definition of the moving particle: it is described
by the vector $\Phi_v$ having a definite nonzero velocity $\vec{v}$. If the
particle were stable, the vector $\Phi_v$ would coincide with $\Psi_p$
at $\vec{p}=\vec{v}m_0\gamma$, $m_0$ being the particle mass. However,
the unstable particle has no definite mass, it is described by a
distribution over masses, see Sect.~2. Therefore, if $\vec{p}$ is definite,
then $\vec{v}$ cannot be definite, see Eq. (\ref{eq12}) below. The exclusion
is the case $\vec{p}=0$ when $\vec{v}$ is also zero.

In the case of unstable particles, whose decay laws can be measured one may
expect that using either $\Phi_v$ or $\Psi_p$ should give only slightly different
results. Indeed, mass distributions of such particles are concentrated in
small regions near average masses $m$, the dimension $\Gamma$ of the regions
being much less than $m$. However, the detailed calculation of $F_v(t)$
presented in Sect.~2 provides instead the unexpected result
$F_v(t)=F_0(t\gamma)$, i.e., contraction instead of dilation: particles
with exact nonzero velocity decay faster than the one at rest.

A simple unstable system is considered in Sect.~4. The oscillating neutrino
may serve as an example. The usual formulae for the neutrino oscillation, e.g.,
see (Bilenky and Pontecorvo, 1978; Bilenky, 2004) are valid when neutrino has
definite momentum $\vec{p}$. The corresponding oscillation is dilated as
compared to the oscillation of the neutrino with lesser momentum.
However, the dilation is not Einsteinian, Eq. (\ref{eq1}). In the case when
neutrino has a definite velocity I obtain another formula for neutrino
oscillation which gives the same contraction as in Sect.~2.

For summary and conclusion see Sect.~5.


\section{DECAY LAW OF MOVING UNSTABLE STATE WITH PRECISE VELOCITY}
\label{s2}

Let us consider a relativistic theory which describes unstable
particles, products of their decay, and the corresponding interactions.
A field theory may be an example. Such a theory must contain
operators of total energy and momentum $\hat{H}$, $\hat{\vec{P}}$
(the generators of time and space translations), total angular
momentum, and generators of Lorentz boosts $\hat{\vec{K}}$.
Usual Dirac's ``instant form" of the theory is implied so that
interaction terms are contained in $\hat{H}$ and $\hat{\vec{K}}$:
$$\hat{H}=\hat{H}_0+\hat{H}_{int},\qquad
\hat{\vec{K}}=\hat{\vec{K}}_0+\hat{\vec{K}}_{int}\,.$$
A simple example is the Lee model of the decay of particle $a$
into two stable particles $b$ and $c$: $a\rightarrow b+c$.
The interaction terms are of the threelinear kind
$\hat{a}\hat{b}^\dag\hat{c}^\dag+H.c.$ (momentum indices of
destruction-creation operators are omitted). The moving particle $a$
is usually described by the eigenvector $\hat{a}_p^\dag\Omega_0$
of the free Hamiltonian $\hat{H}_0$ ($\Omega_0$ is the ``bare" vacuum
and $\hat{a}_p^\dag$ is the creation operator of the particle $a$ with
momentum $\vec{p}$). When $\vec{p}=0$ the vector $\hat{a}_0^\dag\Omega_0$
describes an unstable particle at rest.

Alternatively a moving unstable state may be described by the eigenvector
$\Phi_v=L_v\hat{a}_0^\dag\Omega_0$ of the velocity operator
$\hat{\vec{V}}=\hat{\vec{P}}/\hat{H}$. Here $L_v$ is the Lorentz
transformation from the frame where the velocity of the state
$\hat{a}_0^\dag\Omega_0$ is zero to the frame where the state has
the velocity $v$ (e.g., see Gasiorowicz, 1966). The state
$\hat{a}_p^\dag\Omega_0$ differs from $\Phi_v$:
$\hat{\vec{V}}$ does not commute with $\hat{H}_0$ and, therefore,
$\hat{H}_0$ eigenvector $\hat{a}_p^\dag\Omega_0$ has no definite
velocity.

V. Stefanovich in private communication noted that the state $\Phi_v$
has an admixture of decay particles and, therefore, is not a pure
one-unstable-particle state. Indeed, generators of $L_v$ contain
threelinear (interaction) terms of the kind
$\hat{a}\hat{b}^\dag\hat{c}^\dag$ and consequently
$L_v\hat{a}_0^\dag\Omega_0$ contains a term of the kind
$\hat{b}^\dag\hat{c}^\dag\Omega_0$. This fact does not hinder my
purpose which is to consider an example of an unstable state with
a definite velocity even if it were not a pure one-particle state.
Looking ahead note that Eqs.~(\ref{eq9}) and (\ref{eq16}) below show
that $\Phi_v$ is an really unstable state in the sense that the nondecay
amplitude $(\Phi_v,\Phi_v(t))$, see Eq. (\ref{eq8}), vanishes as
$t\rightarrow\infty$.

To obtain decay laws, the usual solution of the Schroedinger equation
(in the Schroedinger picture) is used: if at $t=0$ the initial state
is $\psi$, then the state $\psi (t)$ at $t>0$ is
$\psi (t)=\exp (-i\hat{H}t)\psi$.

At first consider the initial state $\Phi_0=\hat{a}_0^\dag\Omega_0$. Let
us expand $\Phi_0$ over those eigenvectors $\varphi_\mu$ of $\hat{H}$
which are simultaneously $\hat{\vec{P}}$ or $\hat{\vec{V}}$
eigenvectors with zero eigenvalue (note that $\hat{\vec{P}}$ and
$\hat{\vec{V}}=\hat{\vec{P}}/\hat{H}$ commute with $\hat{H}$). The
corresponding $\hat{H}$ eigenvalues may be called masses and are
denoted by $\mu$: $\hat{H}\varphi_\mu =\mu\varphi_\mu$
\begin{equation}
\label{eq2}
\Phi_0=\int_\mu c(\mu )\varphi_\mu\,,\qquad c(\mu )=(\varphi_\mu ,\Phi_0)\,.
\end{equation}
Then we have
\begin{equation}
\label{eq3}
\Phi_0(t)\equiv\exp (-i\hat{H}t)\Phi_0=\int_\mu c(\mu )\exp (-i\mu t)\varphi_\mu
\end{equation}
and the nondecay (survival) amplitude is
\begin{equation}
\label{eq4}
A_0\equiv (\Phi_0,\Phi_0(t))=\int_\mu |c(\mu )|^2\exp (-i\mu t)\,.
\end{equation}
In Sections 2 and 3 I deal with states for which survival amplitudes vanish as
$t\rightarrow\infty$. This property holds only if the convolution $\int_\mu$
in Eqs. (\ref{eq2})-(\ref{eq4}) is integral over continual $\mu$ values.
Besides, the spectrum of $\hat{H}$ must be bounded from below. So $\int_\mu$ may
be understood as the integral $\int_0^\infty d\mu$.

The vectors $\varphi_\mu$ may be endowed with other indices (e.g., spin ones),
upon which $\hat{H}$ eigenvalues do not depend. I do not write out these
degeneration indices.

Note that the used definition of survival probability $|A_0(t)|^2$
is a particular case of a more general definition, see, e.g., (Exner, 1983).
The latter reduces to $|A_0(t)|^2$ because the used initial states are
eigenvectors of conserving operators $\hat{\vec{P}}$ or $\hat{\vec{V}}$.
In Sect.~4, I shall deal with survival amplitudes that do not vanish as
$t\rightarrow\infty$ but oscillate.

Now let us consider the decay law of the initial state
$\Phi_v=L_v\hat{a}_0^\dag\Omega_0$, see above. Applying the operator $L_v$
to both parts of Eq. (\ref{eq2}) one obtains the expansion of $\Phi_v$
over vectors $L_v\varphi_\mu\equiv\varphi_{v\mu}$:
\begin{equation}
\label{eq5}
\Phi_v=\int d\mu c(\mu )\varphi_{v\mu}\,,\quad
\hat{H}\varphi_{v\mu}=E_{v\mu}\varphi_{v\mu}\,, \quad
\hat{\vec{V}}\varphi_{v\mu}=\vec{v}_{v\mu}\varphi_{v\mu}\,.
\end{equation}
Let us show that $\varphi_{v\mu}$ is $\hat{H}$ eigenvector corresponding
to the eigenvalue $E_{v\mu}=\mu\gamma$, $\gamma =(1-v^2)^{-1/2}$.
Indeed, one has
\begin{equation}
\label{eq6}
L_v^{-1}\hat{H}L_v=(\hat{H}+\vec{v}\hat{\vec{P}})\gamma
\end{equation}
($\hat{H}$ and $\hat{\vec{P}}$ make up a 4-vector). Therefore,
$$
HL_v\varphi_\mu = L_v (\hat{H}+\vec{v}\cdot\widehat{\vec{P}})\gamma\varphi_\mu
=\gamma\mu L_v\varphi_\mu\,.
$$
The equations $\widehat{\vec{P}}\varphi_\mu =0$ and
$\hat{H}\varphi_\mu = \mu\varphi_\mu$ have been used.

Respectively, in place of Eqs. (\ref{eq3}) and (\ref{eq4}) one gets
\begin{eqnarray}
\label{eq7}
\Phi_v(t) &=& \int d\mu\, c(\mu)\,\varphi_{v\mu}\,\exp (-i\mu\gamma t)\,,\\
\label{eq8}
A_v(t)&\equiv& (\Phi_v,\Phi_v(t)) =\int d\mu\, |c(\mu)|^2\,\exp (-i\mu\gamma t)\,.
\end{eqnarray}

\underline{Note.} When calculating Eq. (\ref{eq7}) the orthonormalization
equation $(\varphi_{v\mu_1},\varphi_{v\mu_2})=\delta (\mu_1 - \mu_2)$ is
used, implying unit normalization of $\widehat{\vec{V}}$ eigenvectors.
This is the case if $\widehat{\vec{V}}$ has a discrete spectrum analogously
to the spectrum the momentum has when the system is implied to be in
a large space volume and usual periodicity conditions are imposed (or the
volume opposite boundaries are identified).

Comparing Eq. (\ref{eq8}) with Eq. (\ref{eq4}) one obtains the following
relation of survival amplitudes:
\begin{equation}
\label{eq9}
A_v(t)=A_0(\gamma t)\,.
\end{equation}
The same relation holds for the probabilities $F_v(t)=|A_v(t)|^2$ and
$F_0(t)=|A_0(t)|^2$:
\begin{equation}
\label{eq10}
F_v(t)=F_0(\gamma t)\,.
\end{equation}
So one gets contraction instead of dilation, Eq. (\ref{eq1}), if a moving
unstable state has a definite velocity. Remark that in order to obtain
Eq. (\ref{eq10}) one needs not know $\varphi_\mu$ or $c(\mu )$, see
Eq. (\ref{eq2}); relation (\ref{eq10}) is true for any decay interaction.

In order to discuss this unexpected result I write out the corresponding
survival amplitudes for the state $\psi_p$ with exact momentum.


\section{DECAY LAW OF UNSTABLE STATE WITH PRECISE MOMENTUM}
\label{s3}

I define $\Psi_p$ in analogy with Eq. (\ref{eq5}) by expansion over common
eigenvectors $\psi_{p\mu}$ of the operators $\hat{H}$ and $\hat{\vec{P}}$:
\begin{equation}
\label{eq11}
\Psi_p =\int d\mu c(\mu)\psi_{p\mu}\,,\quad \hat{H}\psi_{p\mu}=E_{p\mu}\psi_{p\mu}\,,
\quad \hat{\vec{P}}\psi_{p\mu}=\vec{p}\psi_{p\mu}\,.
\end{equation}

Here the coefficients $c(\mu )$ do not depend on $p$ just as in Eq. (\ref{eq5}). One
may assume that $\psi_{p\mu}=L_{p\mu}\varphi_\mu$ where $L_{p\mu}$ is the operator of
the Lorentz transformation of the zero velocity state $\varphi_\mu$ into the frame
where the velocity of the state is equal to $\vec{p}/\sqrt{p^2+\mu^2}$, i.e.,
corresponds to the momentum $\vec{p}$. One may verify that $L_{p\mu}\varphi_\mu$
is $\hat{\vec{P}}$ eigenvector with eigenvalue $\vec{p}$. In the same way as before,
see Eq. (\ref{eq6}), one may demonstrate that $\psi_{p\mu}$ is $\hat{H}$ eigenvector
with eigenvalue $E_{p\mu}=\sqrt{p^2+\mu^2}$:
\begin{eqnarray}
&&\hat{H}L_{p\mu}\varphi_\mu=L_{p\mu}L_{p\mu}^{-1}\hat{H}L_{p\mu}\varphi_\mu
=L_{p\mu}(\hat{H}+\vec{v}_{p\mu}\hat{\vec{P}})\gamma_{p\mu}\varphi_\mu\nonumber\\
&&=\mu\gamma_{p\mu}L_{p\mu}\varphi_\mu\,, \qquad
\mu\gamma_{p\mu}=\mu\left[ 1-p^2/(p^2+\mu^2) \right]^{-1/2}=\sqrt{p^2+\mu^2}\,.
\nonumber
\end{eqnarray}
The value $\sqrt{p^2+\mu^2}$ for $E_{p\mu}$ was obtained in a different way in
(Shirokov, 2004).

Let us stress that $\varphi_\mu$ is a stable state (with a definite mass) so
$\psi_{p\mu}$  is also the state with a definite velocity $\vec{v}_{p\mu}$
corresponding to momentum $\vec{p}$. Meanwhile $\Psi_p$ is not eigenstate of
$\hat{\vec{V}}$. Indeed,
\begin{equation}
\label{eq12}
\widehat{\vec{V}}\Psi_p = \int d\mu\, c(\mu)\, \widehat{\vec{P}}/\hat{H}
\psi_{p\mu} = \vec{p} \int d\mu\, c(\mu)\, (p^2+\mu ^2)^{-1/2}\psi_{p\mu} \,.
\end{equation}
So the r.h.s. of (\ref{eq12}) is not proportional to
$\Psi_p = \int d\mu\, c(\mu)\, \psi_{p\mu}$.

Using Eq. (\ref{eq11}) one obtains for survival amplitudes
\begin{eqnarray}
\label{eq13}
A_p(t) &\equiv & \langle \Psi_p\,, e^{-i\hat{H}t}\Psi_p \rangle
= \int d\mu\, |c(\mu)|^2\, \exp (-it\sqrt{p^2+\mu ^2})\,, \\
\label{eq14}
A_0(t) &\equiv & \langle \Psi_0\,, e^{-i\hat{H}t}\Psi_0 \rangle
= \int d\mu\, |c(\mu)|^2\, \exp (-i\mu t)\,.
\end{eqnarray}
Note that when $\vec{p}=0$ the state $\Psi_p(t)=\exp (-i\hat{H}t)\Psi_p$ coincides
with $\Phi_p(t)$, see Eq. (\ref{eq3}).

Now the survival law $A_p(t)$ is not connected with $A_0(t)$ by
such a simple relation as $A_v(t)$ does, see Eq. (\ref{eq9}).
To compare $A_p(t)$ with $A_0(t)$, one has to calculate
$A_p(t)$ and $A_0(t)$ separately. For this purpose one needs to
know $|c(\mu)|^2$. The Breit-Wigner distribution
\begin{equation}
\label{eq15}
|c(\mu)|^2 = \frac{\Gamma}{2\pi}\left[ (\mu -m)^2 +\Gamma ^2/4
\right]^{-1}
\end{equation}
was used in (ESKS). Let us write out approximate expressions
for $A_0(t)$ and $A_p(t)$ which are valid for time not too short and
not too long when the decay laws are exponential (Shirokov, 2004)
\begin{eqnarray}
\label{eq16}
A_0(t) &\cong & \exp (-imt-\frac{1}{2}\Gamma t)\,, \\
\label{eq17}
A_p(t) &\cong & \exp (-im\gamma_m t-\frac{1}{2}\Gamma t/\gamma_m)\,,
\qquad \gamma_m \equiv\sqrt{p^2+m^2}/m\,.
\end{eqnarray}
Here $m$ is the average (or the most probable) mass in the distribution
(\ref{eq15}). It follows from Eqs. (\ref{eq16}) and (\ref{eq17}) that
\begin{equation}
\label{eq18}
|A_p(t)|^2\cong |A_0(t/\gamma_m)|^2 \,,
\end{equation}
i.e. ED holds for survival \underline{probability} of an unstable particle
with precise momentum. The dilation (\ref{eq18}) is to be juxtaposed to the
contraction (\ref{eq10}). As was argued in Introduction, one may expect that
the amplitudes $A_v(t)$ and $A_p(t)$ should not differ appreciably. Let us
show that this expectation is realized in a sense. Note beforehand that
Eqs. (\ref{eq16}) and (\ref{eq9})result in the explicit approximate
expression for $A_v(t)$ when $\vec{v}=\vec{p}/\sqrt{p^2+m^2}$
\begin{equation}
\label{eq19}
A_v(t) \cong  \exp (-im\gamma t-\frac{1}{2}\Gamma t/\gamma)\,.
\end{equation}
Let us compare the exponents $E_p$ and $E_v$ of the corresponding exponentials
in Eqs. (\ref{eq17}) and (\ref{eq19})
\begin{equation}
\label{eq20}
E_p = -imt\gamma_m -\frac{1}{2}\Gamma t/\gamma_m\,,
\qquad E_v = -imt\gamma -\frac{1}{2}\Gamma t\gamma\,,
\end{equation}
assuming that $\gamma_m=\gamma$. As $\Gamma\ll m$ the exponents coincide
in the zero approximation when the terms $\sim \Gamma$ are neglected in
$E_p$ and $E_v$. So in this approximation the corresponding
\underline{amplitudes} $A_p(t)$ and $A_v(t)$ coincide and both satisfy
the contraction property
\begin{equation}
\label{eq21}
A_p(t)\cong A_v(t)\cong A_0(t\gamma)\,.
\end{equation}
However, the main terms of $E_p$ and $E_v$ are purely imaginary and do not
contribute to modules of $A_p(t)$ and $A_v(t)$. It is real parts
$\frac{1}{2}\Gamma t/\gamma_m$ and $\frac{1}{2}\Gamma t\gamma$ that
do contribute and determine the different dependences of
$|A_p|^2$ and $|A_v|^2$ upon $t$, see Eqs. (\ref{eq18}) and (\ref{eq10}).

In the next section I will consider a simple unstable system whose time evolution
is determined by the interference of the main terms defined above (the terms
$\sim\Gamma$ being absent). For this system one may expect the breakdown
of ED (in view of Eq. (\ref{eq21})) even if the system has a precise momentum.


\section{TIME EVOLUTION OF MOVING TWO-MASS STATE}
\label{s4}

Let us consider an unstable system at rest whose state vector $\phi$
is a superposition of two $\hat{H}$ eigenvectors $\varphi_1$
and $\varphi_2$:
\begin{eqnarray}
\label{eq22}
&&\phi =c_1\varphi_1 +c_2\varphi_2\,, \qquad |c_1|^2+|c_2|^2=1\,, \\
\label{eq23}
&&\hat{H}\varphi_1=m_1\varphi_1\,, \qquad
\hat{H}\varphi_2=m_2\varphi_2\,, \qquad m_1\neq m_2\,.
\end{eqnarray}

The system time evolution is described by the survival amplitude
\begin{equation}
\label{eq24}
A_0(t)\equiv (\phi\,,\phi (t))=|c_1|^2 e^{-im_1t}+|c_2|^2 e^{-im_2t}\,.
\end{equation}

The survival amplitudes of the system with nonzero exact velocity $\vec{v}$
and exact momentum $\vec{p}$ are, respectively,
\begin{eqnarray}
\label{eq25}
&&A_v(t)= |c_1|^2 \exp (-im_1t\gamma)+|c_2|^2 \exp (-im_2t\gamma)\,,\\
\label{eq26}
&&A_p(t)= |c_1|^2 \exp (-it\sqrt{p^2+m_1^2})+
|c_2|^2 \exp (-it\sqrt{p^2+m_2^2})\,,
\end{eqnarray}
cf. Eqs. (\ref{eq8}) and (\ref{eq13}).

As examples of such a system one may take electron neutrino, e.g., see
(Bilenky, Pontecorvo, 1978; Bilenky, 2004) and $K_0$-meson
$$
|K_0\rangle = \left( |K_s\rangle +|K_l\rangle \right) /\sqrt{2}\,,
$$
provided that $\Gamma_s=\Gamma_l=0$, e.g., see (Perkins, 1987).
Note that I deal here with the evolution of the unstable states in time,
provided the source and the detector of the states are located in the same
space volume.
In the literature different approaches to neutrino oscillations are
considered: the neutrino source and detector are separated by a distance $R$
and one deals with the oscillatory dependence on $R$, see, e.g.,
(Dolgov, Okun, Rotaev, Schepkin, 2004) and references therein.

In what follows, I let $c_1=c_2=1/\sqrt{2}$. Then
\begin{eqnarray}
\label{eq27}
&&|A_0(t)|^2=\cos ^2\left[ \frac{1}{2}(m_1-m_2)t \right]\,,\\
\label{eq28}
&&|A_v(t)|^2= \cos ^2\left[ \frac{1}{2}(m_1-m_2)t\gamma \right]
=|A_0(t\gamma)|^2\,,\\
\label{eq29}
&&|A_p(t)|^2= \cos ^2\left[ \frac{1}{2}\left(\sqrt{p^2+m_1^2}
-\sqrt{p^2+m_2^2}\right) t \right] =|A_0(t/\tilde{\gamma})|^2\,,\\
&&\tilde{\gamma}=(\sqrt{p^2+m_1^2}+\sqrt{p^2+m_2^2})/(m_1+m_2) \,.
\nonumber
\end{eqnarray}

The oscillatory behavior of these probabilities allows us to use the
two-mass system as the quantum clock. Its unit of time may be defined
as the period of oscillation (the oscillation frequency being equal
to $m_1-m_2$ in the case of $|A_0(t)|^2$), cf. with the definition of
the unit of time provided by radioactive substance, see Introduction.

It follows from (\ref{eq27}) and (\ref{eq29}) that the time evolution
$|A_p(t)|^2$ in the case of exact momentum is dilated as compared to
$|A_0(t)|^2$, but the dilation is not Einsteinian if $m_1\neq m_2$:
$\tilde{\gamma}$ turns into the Lorentz factor only if $m_1\cong m_2$.
In the case of exact velocity we have the same contraction as for unstable
particles, cf. Eqs. (\ref{eq27}) and (\ref{eq28}).


\section{CONCLUSION}
\label{s5}

Relativistic quantum-mechanical derivation of the time evolution of moving
unstable particles was considered in the papers (ESKS). There the state
of the moving particle was defined as the eigenvector $\Psi_p$ of the momentum
operator $\widehat{\vec{P}}$ with eigenvalue $\vec{p}$. It was shown that
then the nondecay law satisfied approximately ED, Eq. (\ref{eq1}). Here in
Sect.~2 the moving particle is described by eigenvector $\Phi_v$ of the
velocity operator $\widehat{\vec{V}}=\widehat{\vec{P}}/\hat{H}$. If the
particle were stable, then $\Phi_v$ would coincide with $\Psi_p$ if
$\vec{p}=\vec{v}m_0\gamma$. The vectors do not coincide in the case of unstable
particle, but one may expect that they should give only slightly different
results. It was shown in Sect.~3 that $\Phi_v$ and $\Psi_p$ give indeed
the same nondecay \underline{amplitude} in the zero approximation. However,
the approximation does not contribute to the corresponding
\underline{probability}, i.e., the nondecay law $F(t)$. As a result, the laws
$F_v(t)$ and $F_p(t)$ turn out to be strongly different: $F_v(t)$ is
contracted as compared to the nondecay law $F_0(t)$ of the particle at
rest, see Eq. (\ref{eq10}), meanwhile $F_p(t)$ is dilated, see Eq. (\ref{eq18}).

Section~4 deals with unstable systems which are simpler than unstable particles.
Oscillating neutrino may be the example. If it has exact momentum, then a
dilation follows, but it is not ED, Eq. (\ref{eq1}). In the case of exact
velocity I obtain the formula which leads to the contraction,
see Eq. (\ref{eq28}).

I conclude that relativistic quantum theory of the time evolution of moving
unstable systems does not ensure ED. The theory allows the possibility of
moving unstable systems whose time evolution breaks ED.

Experiments are known which show that moving mesons have longer lifetimes
than the immovable ones so that ED holds (Bailey, 1977; Farley, 1992).
The theory used here may explain this fact supposing that the experiments
deal with mesons which are in states close to $\Psi_p$. In this case,
the theory approximately gives ED.

\vspace*{1cm}
\noindent
{\Large\bf ACKNOWLEDGMENTS}\\[-0.3cm]

Dr. E. Stefanovich called my attention to that neutrino oscillations
also allow one to verify ED along with $K_0$ oscillation.
I am grateful to him also for useful discussions of this paper.

I thank Dr.~O.~Teryaev for pointing to the paper
(Dolgov et al., 2004).


\newpage
\noindent
{\Large\bf REFERENCES}
\begin{description}
\item Bailey J. et al. (1977). Measurement of relativistic time
dilation for positive and negative muons in a circular orbit.
{\it Nature} {\bf 268}, 301-304.
\item Bilenky, S. M. and Pontecorvo, B. (1978). Lepton mixing and
neutrino oscillations. {\it Physics Report} {\bf 41}, No.~4, 226-261,
Sect.~3.
\item Bilenky, S. M. (2004). Neutrinos: a brief review.
{\it Modern Physics Letters A} {\bf 19}, No.~33, 2451-2477, Sect.~3.
\item Dolgov, A. D., Okun, L. B., Rotaev, M. V. and Schepkin, M. G.
(2004). Oscillation of neutrinos produced by a beam of electrons.
arXiv: hep-ph/0407189.
\item Exner, P. (1983). Representations of the Poincar\'e group
associated with unstable particles. {\it Physical Review D} {\bf 28},
No.~10, 2621-2627.
\item Farley, F. (1992). The CERN (g-2) measurements. {\it Zeitschrift
f\"ur Physik C }{\bf 56}, S88, Sect.~5.
\item Gasiorowicz, S. (1966). {\it Elementary particle physics},
John Wiley, New York, Ch.~4.
\item Khalfin, L. A. (1997). Quantum theory of unstable particles and
relativity. http://www.pdmi.ras.ru/preprint/1997/97-06.html.
\item M{\o}ller, C.(1972). {\it The Theory of Relativity}.
Clarendon Press, Oxford, Ch.~2.6.
\item Perkins, D. (1987). {\it Introduction to High Energy
Physics}. Addison-Wesley Co., Inc.
\item Shirokov, M. I. (2004). Decay law of moving unstable particle.
{\it International Journal of Theoretical Physics} {\bf 43},
No.~6, 1541-1553.
\item Stefanovich, E. (1996). Quantum effects in relativistic
decays. {\it International Journal of Theoretical Physics} {\bf 35},
No.~12, 2539-2554.
\end{description}

\end{document}